# Iteratively Pruned Deep Learning Ensembles for COVID-19 Detection in Chest X-rays

**Sivaramakrishnan Rajaraman**[1]**, Member, IEEE, Jen Siegelman**[2]**, Philip O. Alderson**[3]**, Lucas S. Folio**[4]**, Les R. Folio**[5]**, and Sameer K. Antani**[1]**, Senior Member, IEEE**
[1]Lister Hill National Center for Biomedical Communications, National Library of Medicine, Bethesda, MD 20894 USA
[2]Takeda Pharmaceuticals, Cambridge, MA 02139 USA
[3]School of Medicine, Saint Louis University, St. Louis, MO 63104 USA
[4]Functional and Applied Biomechanics, Clinical Center, National Institutes of Health, and Walt Whitman High School, Bethesda, MD 20817 USA
[5]Radiological and Imaging Sciences, Clinical Center, National Institutes of Health, Bethesda, MD 20894 USA

Corresponding author: Sivaramakrishnan Rajaraman (e-mail: sivaramakrishnan.rajaraman@nih.gov).

This work was supported by the Intramural Research Program of the National Library of Medicine (NLM), and the U.S. National Institutes of Health (NIH).

**ABSTRACT** We demonstrate use of iteratively pruned deep learning model ensembles for detecting pulmonary manifestations of COVID-19 with chest X-rays. This disease is caused by the novel Severe Acute Respiratory Syndrome Coronavirus 2 (SARS-CoV-2) virus, also known as the novel Coronavirus (2019-nCoV). A custom convolutional neural network and a selection of ImageNet pretrained models are trained and evaluated at patient-level on publicly available CXR collections to learn modality-specific feature representations. The learned knowledge is transferred and fine-tuned to improve performance and generalization in the related task of classifying CXRs as normal, showing bacterial pneumonia, or COVID-19-viral abnormalities. The best performing models are iteratively pruned to reduce complexity and improve memory efficiency. The predictions of the best-performing pruned models are combined through different ensemble strategies to improve classification performance. Empirical evaluations demonstrate that the weighted average of the best-performing pruned models significantly improves performance resulting in an accuracy of 99.01% and area under the curve of 0.9972 in detecting COVID-19 findings on CXRs. The combined use of modality-specific knowledge transfer, iterative model pruning, and ensemble learning resulted in improved predictions. We expect that this model can be quickly adopted for COVID-19 screening using chest radiographs.

## I. INTRODUCTION

Novel Coronavirus disease 2019 (COVID-19) is caused by the new Severe Acute Respiratory Syndrome Coronavirus 2 (SARS-CoV-2) that originated in Wuhan in the Hubei province in China and has spread worldwide. The World Health Organization (WHO) declared the outbreak a pandemic on March 11, 2020 [1]. The disease is rapidly affecting worldwide population with statistics quickly falling out of date. As of April 12, 2020, there are over 1.8 million confirmed cases reported globally with over 100,000 reported deaths. Lung disease that causes difficulty in breathing has been reported as an early indicator along with hyperthermia in the COVID-19 infected population [1]. The lung abnormalities caused by non-2019-nCOV viruses are observed as peripheral or hilar and visually similar to, yet often distinct from, viral pneumonia and other bacterial pathogens [2].

Reverse transcription-polymerase chain reaction (RT-PCR) tests are performed to detect the presence of the virus and are considered the gold standard to diagnose COVID-19 infection. However, they are reported to have variable sensitivity and in some geographic regions may not be widely available [3]. While not currently recommended as primary diagnostic tools, chest X-rays (CXRs) and computed tomography (CT) scans have been used to screen for COVID-19 infection and evaluate disease progression in hospital admitted cases [3] [4]. While chest CT offers greater sensitivity to pulmonary disease, there are several challenges to its use. These include the non-portability, the requirement to sanitize the room and equipment between patients followed by a delay of at least an hour [4], the risk of exposing the hospital staff and other patients, and persons under investigation (PUIs) to the virus. Although not as sensitive, portable CXRs are considered as an acceptable alternative [4] since the PUIs can be imaged in more isolated rooms, limiting personnel exposure and because sanitation is much less complex to obtain than with CT.

Automated computer-aided diagnostic (CADx) tools driven by automated artificial intelligence (AI) methods designed to detect and differentiate COVID-19 related thoracic abnormalities should be highly valuable given the heavy burden of infected patients. This is especially important in locations with insufficient CT availability or radiological expertise and CXRs produce fast, high throughput triage such as in a mass casualty [5]. Automated approaches, once validated, have been shown to reduce inter- and intra-observer variability in radiological assessments [6]. Additionally, CADx tools have gained immense significance in clinical medicine by supplementing medical decision making and improving screening and diagnostic accuracy [7]. These tools combine elements of radiological image processing with computer vision for identifying typical disease manifestations and localizing suspicious regions of interest (ROI). At present, recent advances in machine learning, particularly data-driven deep learning (DL) methods using convolutional neural networks (CNNs), have shown promising performance in identifying, classifying, and quantifying disease patterns in medical images. This is particularly true for CT scans and CXRs [7]. These models learn the hierarchical feature representations from medical images to analyze for typical disease manifestations and localize suspicious densities for ROI evaluation [7].

In this study, we highlight the benefits offered through the use of an ensemble of iteratively pruned DL models toward distinguishing CXRs showing COVID-19 pneumonia-related opacities, from bacterial pneumonia, and normals using publicly available CXR collections. Fig. 1 shows the graphical abstract of the proposed study. Fig. 2 shows instances of CXRs being normal, showing bacterial pneumonia, and COVID-19-related pneumonia.

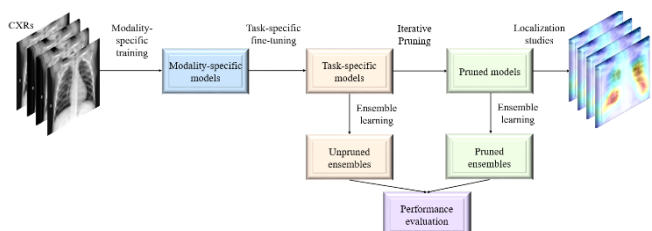

Fig. 1. Graphical abstract of the proposed study.

A custom CNN and a selection of pretrained CNN models are trained on a large-scale selection of CXRs to learn CXR modality-specific feature representations. The learned knowledge then is transferred and fine-tuned to classify the normal and abnormal CXRs. We leverage the benefits of modality-specific knowledge transfer, iterative pruning, and ensemble strategies to reduce model complexity, improve robustness, generalization, and inference capability of the DL model.

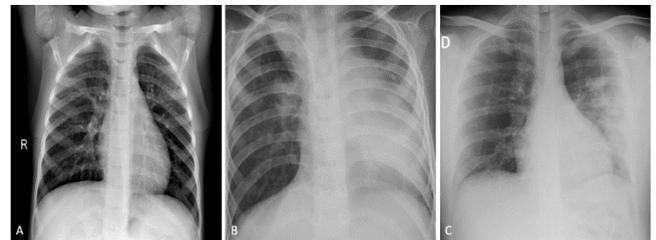

Fig. 2. CXRs showing (A) clear lungs, (B) bacterial pneumonia manifesting as consolidations in the right upper lobe and retro-cardiac left lower lobe, and (C) COVID-19 pneumonia infection manifesting as peripheral opacities in the left lung.

The remainder of the manuscript is organized as follows: Section II discusses prior works. Section III discusses the datasets and methods used toward modality-specific knowledge transfer, iterative pruning, and ensemble learning. Section IV elaborates on the results obtained, and Section V concludes the study with a discussion on the merits and limitations of the proposed approach and future work directions.

## II. PRIOR WORK

*COVID-19 detection*: A study of the literature reveals several AI efforts for COVID-19 screening. The authors of [3] distinguished COVID-19 viral pneumonia manifestations from that of other viral pneumonia on chest CT scans with high specificity. It was observed that COVID-19 pneumonia was found to be peripherally distributed with ground glass opacities (GGO) and vascular thickening. The authors of [8] established a publicly available collection of 275 CT scans showing COVID-19 pneumonia manifestations and trained a deep CNN to achieve 0.85 F-score in classifying CTs as normal or showing COVID-19 pneumonia-related opacities. The authors of [9] used a customized CNN and pretrained AlexNet model to classify CXRs as normal or showing COVID-19 pneumonia with 94.1% and 98% accuracy respectively. The authors of [10] used a ResNet-50 [11] CNN to classify normal, pneumonia, and COVID-19 viral pneumonia manifestations in CXRs and achieved an accuracy of 98.18 % and F-score of 98.19. CXRs are also commonly analyzed to diagnose and differentiate other types of pneumonia including bacterial and non-COVID-19 viral pneumonia [2]. The authors of [12] proposed a custom CNN model that was designed by combining manual design prototyping with a machine-driven designing approach to classify CXRs as normal or showing non-COVID-19 or COVID-19 pneumonia-related opacities with 92.4% accuracy.

*Modality-specific knowledge transfer*: With limited amounts of COVID-19 pneumonia CXR data, traditional transfer learning strategies offer promise [13] where the learned feature representations are fine-tuned to improve performance. However, unique challenges posed in the appearance of medical images [6] including high inter-class similarity and low intra-class variance lead to model bias and overfitting resulting in reduced performance and generalization. These issues can be alleviated through modality-specific knowledge transfer by retraining CNN models on a large CXR image collection to learn modality-specific feature representations. Modality-specific model knowledge transfer [14] and ensembles [15] have demonstrated superior disease ROI localization compared to individual constituent models.

*Model pruning*: To alleviate burdens from computing resources, DL models can be pruned to reduce the inference cost and facilitate deployment in low-resource conditions with no loss or even improvement in performance. Reed [16] performed a neural model pruning to decrease computational complexity. Hassibi & Stork [17] deleted network parameters by leveraging the second derivative term in the Taylor series and improved model generalization. The authors of [18] found that the earlier layers in the neural networks have low activations that can effectively be excluded from the network without affecting the model performance. They proposed an iterative optimization method to gradually eliminate the neurons with the least activations toward reducing the memory and power requirements and promoting faster model inference. When applied to medical imaging, the authors of [19] proposed a genetic algorithm-based pathway evolution strategy to prune DL models. This resulted in a 34% reduction in the network parameters and improved the mass classification performance in breast mammograms. A systematic weight pruning strategy [20] was used to prune a YOLO-model [21] based pneumonia detector for classifying CXRs as normal or showing pneumonia-like manifestations using the Radiological Society of North America (RSNA) [22] CXR collection. However, there is room for further research in this area.

*Ensemble classification*: CNNs are non-linear models that learn complex relationships from the data through error backpropagation and stochastic optimization, making them highly sensitive to random weight initializations and the statistical noise present in the training data. These issues can be alleviated by ensemble learning by training multiple models and combining their predictions where an individual model's weaknesses are offset by the predictions of other models. Combined predictions are shown to be superior to individual models [23]. There are several ensemble strategies reported in the literature including max voting, simple and weighted averaging, stacking, boosting, blending, and others that are shown to minimize the variance error and improve generalization and performance of CNN models. Applied to CXRs, the authors of [7], [14], and [24] leveraged the use of an ensemble of CNN models toward improving TB detection in CXRs. An averaging ensemble of pretrained CNNs was used by the authors of [25] toward improving cardiomegaly detection using CXRs.

## III. MATERIALS AND METHODS

### A. DATA COLLECTION AND PREPROCESSING

Table 1 shows the distribution of CXRs across different categories. We used the following four publicly available CXR collections in this retrospective analysis:

TABLE 1
DATASET CHARACTERISTICS. NUMERATOR AND DENOMINATOR DENOTES THE NUMBER OF TRAIN AND TEST DATA RESPECTIVELY (N = NORMAL, UP=PNEUMONIA OF UNKNOWN TYPE, BP= BACTERIAL (PROVEN) PNEUMONIA, CP = COVID-19 PNEUMONIA)

| Dataset | N | UP | BP | CP |
|---------|---|----|----|----|
| A | 1349/234 | 0 | 2538/242 | 0 |
| B | 5412/600 | 5412/600 | 0 | 0 |
| C | 0 | 0 | 0 | 121/13 |
| D | 0 | 0 | 0 | 165/14 |

A) Pediatric CXR dataset [2]: The authors collected from Guangzhou Women and Children's Medical Center in Guangzhou, China, the anterior-posterior (AP) CXRs of children from 1 to 5 years of age, showing normal lungs, bacterial pneumonia, and non-COVID-19 viral pneumonia. Expert radiologists curated the CXR collection to remove low-quality chest radiographs.

B) RSNA CXR dataset [22]: This multi-expert curated dataset includes images from the National Institutes of Health (NIH) CXR-14 dataset [26]. The dataset was released for the Kaggle pneumonia detection challenge, organized jointly by RSNA and NIH. The collection includes normal CXRs and abnormal images with non-pneumonia and pneumonia-like opacities. The images are made available at 1024 × 1024 pixel resolution in DICOM format.

C) Twitter COVID-19 CXR dataset: A cardiothoracic radiologist from Spain made available a collection of 134 CXRs with 2K×2K pixel resolution in JFIF format via Twitter of SARS-CoV-2 positive subjects. (https://twitter.com/ChestImaging)

D) Montreal COVID-19 CXR dataset [27]: A publicly available periodically updated GitHub repository that includes COVID-19 CXR cases and other pulmonary viral disease manifestations in AP, posterior-anterior (PA), and AP-Supine views. As of April 7, 2020, the repository had 179 CXRs showing COVID-19 pneumonia-related opacities.

We performed patient-level splits of these CXR collections to allocate 90% for training and 10% for testing at different stages of learning discussed in this study. We randomly allocated 10% of the training data to validate the DL models. The ground truth (GT) for the test set, comprising of 27 CXRs showing COVID-19 pneumonia-related opacities is set by the verification of publicly identified cases from expert radiologists who annotated the test set.

### B. LUNG ROI SEGMENTATION

While mild COVID-19 cases mimic common upper respiratory viral infections, advanced disease results in respiratory dysfunction and is the principal cause for triggering mortality. In developing DL solutions for detecting the disease, it is important to guard them against irrelevant features that could severely affect reliable decision-making. For this study, we performed U-Net based semantic segmentation [28] to segment the lung pixels from the background. We used a U-Net with Gaussian dropout layers [29] added to the U-Net encoder. A dropout ratio of 0.2 was empirically determined and used in this study. Fig. 3 illustrates the segmentation steps performed in this study.

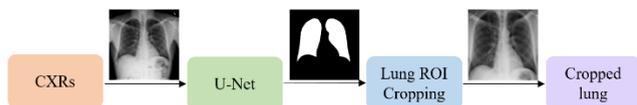

Fig. 3. The segmentation approach showing U-Net based mask generation and Lung ROI cropping.

We used a collection of CXRs with lung masks from [30] to train the U-Net model to generate lung masks of 256×256 pixel resolution for the aforementioned datasets. We used model checkpoints to monitor its performance and stored only the best model weights to generate the final lung masks. These masks then are superimposed on the CXR images to crop them as a bounding box containing the lung pixels. The cropped lungs are resized to 256×256 pixel resolution. The lung crops are further preprocessed by performing pixel rescaling, median filtering for noise removal and edge preservation, normalization for mean, and standardization for identical feature distribution. The preprocessed lung crops are used for model training and evaluation at different stages of learning discussed in this study.

### C. MODELS AND COMPUTATIONAL RESOURCES

We evaluated the performance of a customized CNN and a selection of ImageNet pretrained CNN models, viz., a) VGG-16 [31], b) VGG-19 [31], c) Inception-V3 [32], d) Xception [33], e) InceptionResNet-V2 [32]; f) MobileNet-V2 [34], g) DenseNet-201 [35], and h) NasNet-mobile [36].

Our customized CNN is a linear stack of strided separable convolution layers, global average pooling (GAP), and a dense layer with Softmax activation. Fig. 4 shows the architecture of the custom CNN used in this study. We used Dropout to reduce issues due to model overfitting by providing restricted regularization and improving generalization by reducing the model sensitivity to the specifics of the training input [29]. We used strided convolutions that were shown to improve performance on several visual recognition benchmarks, compared to max-pooling layers [37]. Separable convolutions were used to reduce model parameters [33] and improve performance compared to conventional convolution operations. The number of separable convolutional filters are initialized to 32 and increased by a factor of two in the successive convolutional layers. We used 5×5 filters and a stride length of 2 in all convolutional layers. We added a GAP layer to average the spatial feature dimensions that are fed into the final dense layer with Softmax activation.

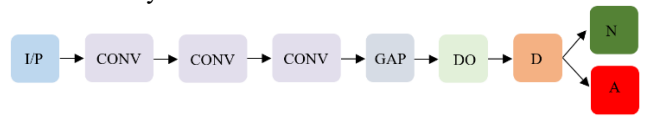

Fig. 4. Architecture of the customized CNN model. (I/P = Input, CONV = Convolution, GAP = Global average pooling, DO = Dropout, D = Dense with Softmax activation, N = Normal predictions, A = Abnormal Predictions).

We used the Talos optimization package [38] to optimize the parameters and hyperparameters of the customized CNN that include a) dropout ratio, b) optimizer and c) non-linear activation function. The model is trained and evaluated with the optimal parameters to classify the CXRs to their respective categories.

We instantiated the pretrained CNN with their ImageNet weights and truncated them at the fully-connected layers. The following layers are added to the truncated model: (a) zero-padding, (b) a strided separable convolutional layer with 5×5 filters and 1024 feature maps, (c) GAP layer, (d) Dropout layer with an empirically determined dropout ratio of 0.5, and (e) final dense layer with Softmax activation. Fig. 5 shows the customized architecture of the pretrained models used in this study.

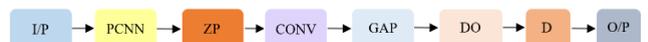

Fig. 5. Architecture of the pretrained CNNs. (I/P = Input, PCNN = truncated model, ZP = Zero-padding, CONV = Convolution, GAP = Global Average Pooling, DO = Dropout, D=Dense with Softmax activation, O/P = Output).

We optimized the following hyperparameters of the pretrained CNNs using a randomized grid search method [39]: (a) momentum, (b) L2-regularization, and (c) initial learning rate of the Stochastic Gradient Descent (SGD) optimizer. The search ranges were initialized to [0.85 0.99], [1e−10 1e−3], and [1e−9 1e−2] and for the momentum, L2-regularization, and the initial learning rate respectively. The pretrained CNNs were retrained with smaller weight updates to improve generalization and categorize the CXRs to their respective classes. Class weights were used during model training to penalize the overrepresented classes to prevent overfitting and improve performance [40]. We used model checkpoints to store the best model weights for further analysis.

### D. MODALITY-SPECIFIC TRANSFER LEARNING AND FINE-TUNING

We performed modality-specific transfer learning where the customized CNN and ImageNet pretrained models are retrained on the RSNA CXR collection to learn CXR modality-specific features and classify the CXRs into normal and abnormal categories. The RSNA CXR collection includes normal CXRs and abnormal images containing pneumonia-related opacities. In this way, the weight layers are made specific to the CXR modality through learning the features of normal and abnormal lungs. The learned knowledge is transferred and fine-tuned to a related task of classifying CXRs that are pooled from pediatric, Twitter COVID-19, and Montreal COVID-19 CXR collections, respectively, as normal, or showing bacterial pneumonia, or COVID-19 pneumonia-related opacities, to improve classification performance.

The top-3 performing modality-specific CNNs are instantiated and truncated at their deepest convolutional layer and added with the following layers: (a) zero-padding, (b) a strided separable convolutional layer with 5×5 filters and 1024 feature maps, (c) GAP layer, (d) Dropout layer and (e) final dense layer with Softmax activation. The modified models are fine-tuned to classify CXRs as being normal or showing bacterial pneumonia or COVID-19 viral pneumonia. Class weights were used during model training to award higher weights to the under-represented class to reduce issues due to class imbalance and improve generalization and performance. Fine-tuning is performed through SGD optimization and model checkpoints were used to store the best weights for further analysis.

### E. ITERATIVE MODEL PRUNING

We iteratively pruned the fine-tuned models to find the optimal number of neurons in the convolutional layers to reduce model complexity with no loss in performance. We gradually eliminated the neurons with fewer activations at each time step through iterative pruning and model retraining. We used the average percentage of zeros (APoZ) [18], the percentage of zero neuron activations observed with the validation dataset, as the measure to rank the neurons in each convolutional layer. We iteratively pruned a percentage of neurons with the highest APoZ from each layer at each time step and retrained the pruned model. The process is repeated until the maximum percentage of pruning is achieved. The best-pruned model is then selected from the collection of iteratively pruned models based on their performance with the test set. The retrained pruned model is expected to achieve similar or better performance than the unpruned models with reduced model complexity and computational requirements. The algorithm for iterative pruning performed in this study is described below:

---
Algorithm 1 – Iterative Pruning

---

**Input:** $B = \{(x_i, y_i) \mid x_i \in X, y_i \in Y\}$, pruning percentage (P), maximum pruning percentage (M)

1. Train and evaluate the base models on $B$ and store the best model weights
2. **while** *percent pruned (PP) <= M* **do**

   a. Calculate the number of filters in each convolutional layer
   b. Identify and delete $P$ percentage of filters in each convolutional layer with the highest average percentage of zeros
   c. Retrain and evaluate the pruned model on $B$ and store the best-pruned weights
   d. $PP$ += $P$
   e. Incrementally prune the network, retraining it each time and save the pruned model
   **end while**

**Return:** $M+1$ number of pruned models

### F. LEARNING ITERATIVELY PRUNED ENSEMBLES

The best performing pruned models are selected to construct the ensemble to improve prediction performance and generalization as compared to any individual constituent model. We used several ensemble strategies including max voting, averaging, weighted averaging, and stacking to combine the predictions of the pruned models toward classifying CXRs as normal or showing bacterial or COVID-19 viral pneumonia-related opacities. For the stacking ensemble, we used a neural network-based meta-learner that learns to optimally combine the predictions of the individual pruned models. The meta-learner consisting of a single hidden layer with nine neurons is trained to interpret the multi-class input from the top-3 pruned models and a final dense layer outputs the predictions to categorize the CXRs to their respective classes.

### G. VISUALIZATION STUDIES

Visualizing the learned behavior of the DL models is a debated topic, particularly in medical visual recognition tasks. There are several visualization strategies reported in the literature that include (a) visualizing the overall network structure and (b) gradient-based visualization that performs gradient manipulation during network training. Gradient-weighted class activation mapping (Grad-CAM) is a gradient-based visualization method that computes the scores for a given image category concerning the feature maps of the deepest convolutional layer in a trained model [41]. The gradients that are flowing backward are pooled globally to measure the importance of the weights in the decision-making process. In this study, we verified the learned behavior of the pruned models by comparing salient ROI with consensus GT annotations from experienced radiologists.

### H. STATISTICAL ANALYSES

We analyzed the model's performance for statistical significance at different stages of learning. We used confidence intervals (CI) as the measure to analyze the skill of the CNN models. A shorter CI infers a smaller margin of error or a relatively precise estimate while a larger CI allows more margin for error and therefore results in reduced precision [42]. We computed the 95% CI values for the AUC at different learning stages to explain the models' predictive performance. The CI values are computed to be the Clopper–Pearson exact interval that corresponds to the separate 2-sided interval with individual coverage probabilities of $(0.95)^{1/2}$. We used StatsModels version 0.11.0 to compute CI measures. The codes associated with this study are made available at https://github.com/sivaramakrishnan-rajaraman/Iteratively-pruned-model-ensembles-for-COVID-19-detection-in-CXRs.

## IV. RESULTS AND DISCUSSION

The optimal values for the parameters and hyperparameters obtained for the customized and pretrained CNNs with the Talos optimization tool and randomized grid search, respectively, are shown in Table 2.

TABLE 2
OPTIMAL VALUES FOR THE PARAMETERS AND HYPERPARAMETERS FOR THE CUSTOM AND PRETRAINED MODELS OBTAINED THROUGH OPTIMIZATION TOOLS (M = MOMENTUM, ILR = INITIAL LEARNING RATE, L2 = L2-WEIGHT DECAY, AND D = DROPOUT RATIO)

| Models | Optimal values | | | | | |
|---|---|---|---|---|---|---|
| | M | ILR | L2 | D | Optimizer | Activation |
| Custom | - | - | - | 0.5 | SGD | ReLU |
| Pretrained | 0.95 | 1e-3 | 1e-6 | - | - | - |

Table 3 shows the performance achieved through modality-specific knowledge transfer, by the customized and pretrained CNNs using the RSNA CXR dataset.

TABLE 3
PERFORMANCE METRICS ACHIEVED DURING MODALITY-SPECIFIC TRANSFER LEARNING USING THE RSNA CXR DATASET (ACC. = ACCURACY; SENS. = SENSITIVITY, PREC. = PRECISION, F = F-SCORE, MCC = MATTHEWS CORRELATION COEFFICIENT, AND PARAM. = TRAINABLE PARAMETERS). THE VALUES IN SQUARE BRACKETS SHOW THE 95% CI THAT ARE COMPUTED TO BE THE CLOPPER–PEARSON EXACT INTERVAL CORRESPONDING TO THE SEPARATE 2-SIDED INTERVAL WITH INDIVIDUAL COVERAGE PROBABILITIES OF $(0.95)^{1/2}$.

| Models | Acc. | AUC | Sens. | Prec. | F | MCC | Param. |
|---|---|---|---|---|---|---|---|
| Custom | 0.9467 | 0.9842 [0.9691 0.9993] | 0.9434 | 0.9497 | 0.9465 | 0.8934 | 47885 |
| VGG-16 | 0.9750 | 0.9919 [0.9817 1.0] | 0.9684 | 0.9812 | 0.9752 | 0.9501 | 19436354 |
| VGG-19 | 0.9717 | 0.9923 [0.9817 1.0] | 0.9717 | 0.9749 | 0.9716 | 0.9434 | 24746050 |
| Inception-V3 | 0.9683 | 0.9922 [0.9815 1.0] | 0.9584 | 0.9879 | 0.9685 | 0.9375 | 40645794 |
| Xception | 0.9667 | 0.9896 [0.9773 1.0] | 0.9517 | 0.9811 | 0.9661 | 0.9332 | 39684394 |
| DenseNet-201 | 0.9683 | 0.991 [0.9795 1.0] | 0.9634 | 0.9782 | 0.9677 | 0.9377 | 16394114 |
| MobileNet-V2 | 0.9675 | 0.9903 [0.9784 1.0] | 0.955 | 0.9715 | 0.9674 | 0.9351 | 12647234 |
| NASNet-mobile | 0.9600 | 0.9882 [0.9751 1.0] | 0.9584 | 0.9616 | 0.96 | 0.9201 | 13968148 |
| InceptionResNet-V2 | 0.9625 | 0.9887 [0.9759 1.0] | 0.9417 | 0.9827 | 0.9618 | 0.9259 | 68435042 |

It can be observed that the VGG-16, VGG-19, and Inception-V3 models were more accurate than the other models under study. The aforementioned models demonstrated promising AUC values with a shorter CI and hence a smaller margin of error, thereby offering precise estimates compared to the other models. This is because the architecture depths of the VGG and Inception-V3 models are optimal to learn the hierarchical representations of features from the CXR data and classify them into normal and pneumonia classes. Considering the F-score and MCC that give a balanced measure of precision and recall, the aforementioned models delivered performance that was superior to the other models.

The top-3 performing modality-specific knowledge transfer models (VGG-16, VGG-19, and Inception-V3) are instantiated with their modality-specific weights and truncated at their fully connected layers and appended with the task-specific heads. Table 4 shows the performance achieved by the task-specific models toward the following classification tasks: (a) binary classification to classify CXRs as normal or COVID-19 pneumonia and (b) multi-class classification to classify CXRs as normal or as showing bacterial pneumonia or COVID-19 pneumonia.

TABLE 4
PERFORMANCE METRICS ACHIEVED BY THE TOP-3 MODALITY-SPECIFIC KNOWLEDGE TRANSFER MODELS ON THE TARGET TASKS

| Task | Models | Acc. | AUC | Sens. | Prec. | F | MCC | Param. |
|---|---|---|---|---|---|---|---|---|
| Normal vs. COVID-19 | VGG-16 | 1 | 1 [1 1] | 1 | 1 | 1 | 1 | 19436354 |
| | VGG-19 | 1 | 1 [1 1] | 1 | 1 | 1 | 1 | 24746050 |
| | Inception-V3 | 1 | 1 [1 1] | 1 | 1 | 1 | 1 | 40645794 |
| Normal vs. Bacterial vs. COVID-19 | VGG-16 | 0.9702 | 0.9977 [0.9935 1.0] | 0.9702 | 0.9709 | 0.9702 | 0.9465 | 19437379 |
| | VGG-19 | 0.9563 | 0.9936 [0.9866 1.0] | 0.9563 | 0.9588 | 0.9563 | 0.9226 | 24747075 |
| | Inception-V3 | **0.9742** | **0.9969 [0.9921 1.0]** | **0.9742** | **0.9746** | **0.9742** | **0.9534** | 40646819 |

*Bold values stand for the model with a statistically significant better performance than the other models.

It can be observed that for the binary classification task, all the models are 100% accurate, however, VGG-16 has the least number of trainable parameters. For multi-class classification, it can be observed that the Inception-V3 model was more accurate with a shorter CI for the AUC metric, signifying that it has the least margin for error and hence provides a more precise estimate. Considering F-score and MCC, the Inception-V3 model delivered superior performance compared to VGG-16 and VGG-19 models.

For the multi-class classification task, the predictions of the task-specific models (VGG-16, VGG-19, and Inception-

V3) are combined through several ensemble methods including max voting, simple averaging, weighted averaging, and model stacking. We didn't perform ensemble learning for the binary classification task since the individual models are 100% accurate in classifying CXRs as normal or showing COVID-19 pneumonia-related opacities. Table 5 shows the performance achieved for the multi-class classification with different ensemble strategies. It can be observed that a simple average of the models' predictions is more accurate with a shorter CI for the AUC metric, signifying a smaller margin of error and therefore, higher precision, compared to other ensemble methods. Considering the F-score and MCC, the averaging ensemble outperformed other ensemble strategies in classifying CXRs as normal, or as showing bacterial pneumonia or COVID-19 viral pneumonia.

For the multi-class classification task, we iteratively pruned the task-specific models (VGG-16, VGG-19, and Inception-V3) by removing 2% of the neurons with the highest APoZ in each convolutional layer at a given time step and retrained the pruned model to evaluate its performance on the validation set. We used model checkpoints to store the best-pruned model that gave a superior performance with the validation set. The process is repeated until the maximum pruning percentage of 50% is reached. We then evaluated the performance of all the pruned models on the test set. The pruned model that achieved superior performance with the test set is used for further analysis.

TABLE 5
PERFORMANCE METRICS ACHIEVED BY THE UNPRUNED MODELS THROUGH DIFFERENT ENSEMBLE STRATEGIES FOR THE MULTICLASS CLASSIFICATION TASK

| Method | Acc. | AUC | Sens. | Prec. | F | MCC |
|---|---|---|---|---|---|---|
| Majority Voting | 0.9742 | 0.9807 [0.9686 0.9928] | 0.9742 | 0.9748 | 0.9742 | 0.9537 |
| Averaging | **0.9782** | **0.9969 [0.992 1.0]** | **0.9782** | **0.9786** | **0.9782** | **0.9607** |
| Weighted Averaging | 0.9762 | 0.9968 [0.9918 1.0] | 0.9762 | 0.9767 | 0.9762 | 0.9572 |
| Stacking | 0.9663 | 0.9865 [0.9764 0.9966] | 0.9663 | 0.968 | 0.9662 | 0.9402 |

*Bold values stand for the method with a statistically significant better performance than the other ensemble methods.

Table 6 shows a comparison of the performance achieved by the pruned models to that of the baseline, unpruned task-specific models shown in Table 4. It can be observed that the pruned models are more accurate than their unpruned counterparts. Considering the F-score and MCC metrics, the pruned models are found to deliver superior performance than the unpruned models. It is interesting to note that the performance improvement is achieved with a significant reduction in the number of parameters. As can be seen, the number of parameters in the pruned VGG-16 model reduced by 46.03% compared to its unpruned counterpart. Similarly, the number of trainable parameters reduced by 16.13% and 36.1% for the pruned VGG-19 and Inception-V3 models, respectively with the added benefit of performance improvement in terms of accuracy, F-score, and MCC metrics, compared to their unpruned counterparts.

TABLE 6
PERFORMANCE METRICS ACHIEVED BY THE BEST ITERATIVELY PRUNED MODELS AND COMPARED WITH THE BASELINE UNPRUNED MODELS FROM TABLE 4 (U-UNPRUNED AND P-PRUNED)

| Models | Acc. | AUC | Sens. | Prec. | F | MCC | Param. | % Reduction |
|---|---|---|---|---|---|---|---|---|
| VGG-16-U | 0.9702 | 0.9977 [0.9935 1.0] | 0.9702 | 0.9709 | 0.9702 | 0.9465 | 19437379 | |
| VGG-16-P | 0.9722 | 0.9938 [0.9869 1.0] | 0.9722 | 0.9725 | 0.9722 | 0.9498 | 10490921 | 46.03 |
| VGG-19-U | 0.9563 | 0.9936 [0.9866 1.0] | 0.9563 | 0.9588 | 0.9563 | 0.9226 | 24747075 | |
| VGG-19-P | 0.9762 | 0.9972 [0.9925 1.0] | 0.9762 | 0.9767 | 0.9762 | 0.9572 | 20756001 | 16.13 |
| Inception-V3-U | 0.9742 | 0.9969 [0.9992 1.0] | 0.9742 | 0.9746 | 0.9742 | 0.9534 | 40646819 | |
| Inception-V3-P | 0.9841 | 0.9962 [0.9908 1.0] | 0.9841 | 0.9841 | 0.9841 | 0.9712 | 25994510 | 36.10 |

Fig. 6 shows the results of performing Grad-CAM visualizations to localize the salient ROIs used by the different pruned models to classify a sample test CXR into the COVID-19 viral pneumonia category. The visualizations are compared with consensus GT annotations provided by the expert radiologists. The predictions of the pruned models are decoded for the test sample. Two-dimensional heat maps are generated in bright red, which corresponds to the pixels carrying higher importance and hence weights for categorizing the test sample to COVID-19 pneumonia infected category. Distinct color transitions are observed for varying ranges of pixel importance toward making the predictions. Salient ROIs are localized by superimposing the heat maps on the input sample CXR. It is observed that the pruned models precisely localize the salient ROI. This underscores the fact that the pruned models have learned the implicit rules that generalize well and conform to the experts' knowledge about the problem.

Table 7 shows a comparison of the performance metrics achieved with the different ensemble strategies for the unpruned and pruned models toward classifying the CXRs as normal or showing bacterial pneumonia, or COVID-19 viral pneumonia.

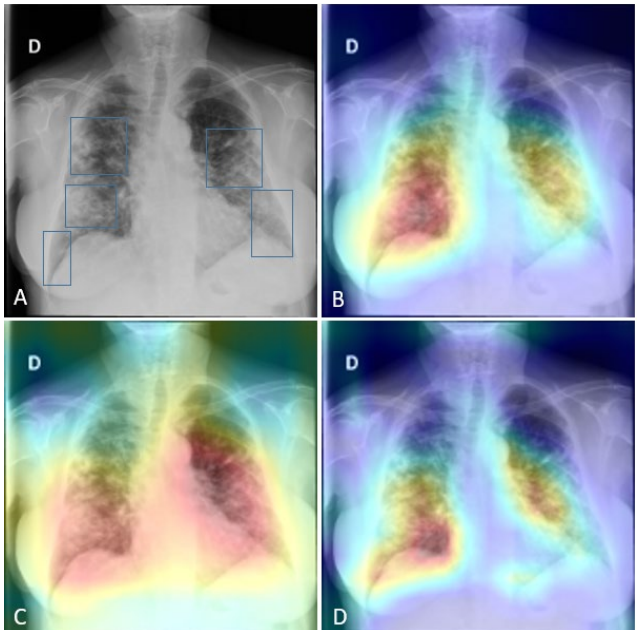

Fig. 6. Grad-CAM Visualizations showing salient ROI detection by different pruned models. (A) CXR showing COVID-19 viral pneumonia-related opacities with GT annotations, (B) VGG-16 pruned model, (C) VGG-19 pruned model, and (D) Inception-V3 pruned model. Bright red corresponds to the pixels carrying higher importance and hence weights for categorizing the test sample to the COVID-19 viral pneumonia category.

TABLE 7
COMPARING THE PERFORMANCE METRICS ACHIEVED WITH THE PRUNED AND UNPRUNED MODEL ENSEMBLES FROM TABLE 4

| Method | Method | Acc. | AUC | Sens. | Prec. | F | MCC |
|---|---|---|---|---|---|---|---|
| Majority Voting | Unpruned | 0.9742 | 0.9807 [0.9686 0.9928] | 0.9742 | 0.9748 | 0.9742 | 0.9537 |
|  | Pruned | 0.9821 | 0.9866 [0.9765 0.9967] | 0.9821 | 0.9822 | 0.9821 | 0.9676 |
| Averaging | Unpruned | 0.9782 | 0.9969 [0.992 1.0] | 0.9782 | 0.9786 | 0.9782 | 0.9607 |
|  | Pruned | 0.9821 | 0.9969 [0.992 1.0] | 0.9821 | 0.9823 | 0.9821 | 0.9677 |
| Weighted Averaging | Unpruned | 0.9762 | 0.9968 [0.9918 1.0] | 0.9762 | 0.9767 | 0.9762 | 0.9572 |
|  | Pruned | **0.9901** | **0.9972 [0.9925 1.0]** | **0.9901** | **0.9901** | **0.9901** | **0.9820** |
| Stacking | Unpruned | 0.9663 | 0.9865 [0.9764 0.9966] | 0.9663 | 0.968 | 0.9662 | 0.9402 |
|  | Pruned | 0.9712 | 0.9876 [0.9779 0.9973] | 0.9712 | 0.9711 | 0.9712 | 0.9473 |

*Bold values stand for the model with a statistically significant better performance than the other models.

While performing weighted averaging ensemble for both unpruned and pruned models, the predictions are awarded the importance based on their F-score and MCC measures that offer a balanced measure of precision and sensitivity. From Table 6, it can be observed that the pruned and unpruned Inception-V3 model delivered superior performance, followed by VGG-19 and VGG-16 models. In this regard, we assigned weights of 0.5, 0.3, and 0.2 to the predictions of Inception-V3, VGG-19, and VGG-16 models, respectively. It can be observed that the weighted averaging ensemble of the predictions of the pruned models delivered superior performance in all aspects. Fig. 7 and Fig. 8 shows the confusion matrix and AUC curves, respectively, obtained with the weighted-averaging pruned ensemble.

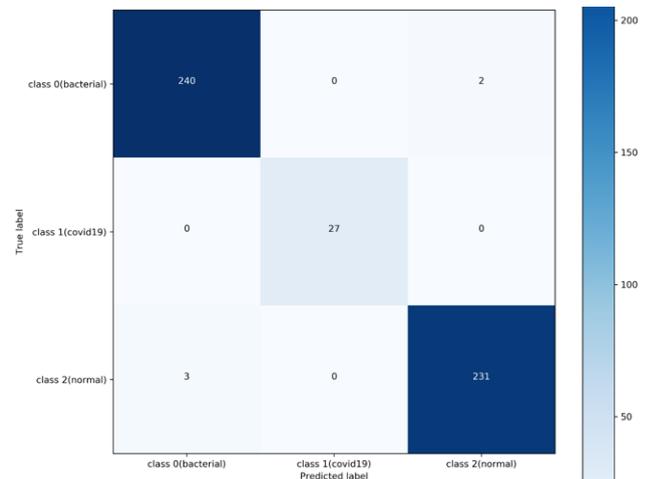

Fig. 7. Confusion matrix obtained with the weighted-average pruned ensemble.

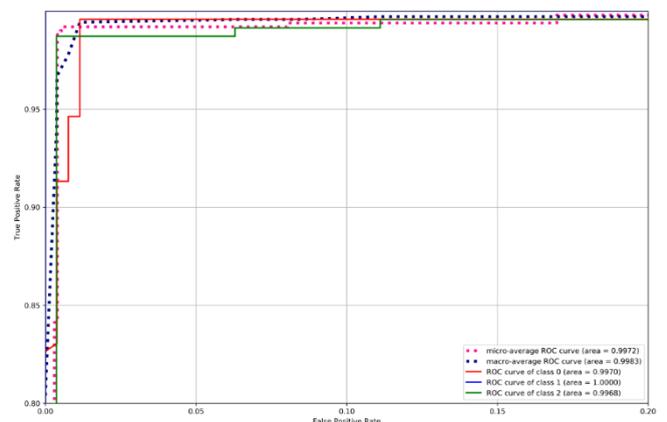

Fig. 8. ROC curves showing micro/macro-averaged and class-specific AUC obtained with the weighted-average pruned ensemble.

The 95% CI for the AUC metric has the shortest error margin with a more precise estimate than that obtained with the other ensemble methods. Considering the F-score and MCC, the weighted averaging ensemble outperformed the other ensemble strategies in classifying CXRs as normal, bacterial pneumonia, or COVID-19 viral pneumonia.

## V. CONCLUSION
The COVID-19 pandemic has had an enormously negative impact on population health and national economies worldwide. Early diagnosis has often been suboptimal and serological tests have not been widely available. The opportunity to utilize CXRs as part of the diagnostic approach could add an important and nearly universally available tool to the battle against COVID-19 or other respiratory viruses that might emerge in the future. In the current study, we demonstrate that this can be done by applying ensemble DL to findings seen in CXRs.

Modality-specific transfer learning performed with a large-scale CXR collection with a diversified data distribution helped in learning CXR modality-specific features. The learned feature representations served as a good weight initialization and improved model adaptation and generalization compared to ImageNet pretrained weights, when transferred and fine-tuned for a related CXR classification task.

Iterative pruning of the task-specific models and selection of the best performing pruned model not only improved prediction performance on the test data but also significantly reduced the number of trainable parameters. This is because there are redundant network parameters (neurons) in a deep model that do not contribute to improving the prediction performance. If these neurons with lesser activations can be identified and removed, it results in a faster and smaller model with similar or improved performance than the unpruned models. This would facilitate deploying these models on browsers and mobile devices.

We further improved the performance by constructing ensembles of the pruned models. By empirically evaluating the performance of the pruned models and awarding weights based on their predictions, we observed that the weighted averaging ensemble of the pruned models outperformed the other ensemble methods.

We performed visualization studies to validate the pruned model localization performance and found that the pruned models precisely localized the salient ROI used in categorizing the input CXRs to their expected categories.

We observe that combined use of CXR modality-specific knowledge transfer, iterative model pruning, and ensemble learning reduced prediction variance, model complexity, promoted faster inference, performance, and generalization. However, the success of this approach is controlled by two broad factors: (i) dataset size and inherent variability, and (ii) computational resources needed for successful deployment and use. With dataset size, we specifically refer to the minimum number of topically relevant images, in this case, CXRs with viral pneumonia that are distinct from bacterial and normal images, that are needed to build confidence into the ensemble. With computational resources, we recognize the training time and memory constraints required for practicable deployment. However, low-cost GPU solutions, high-performance computing (HPC), and cloud technology would address the feasibility in this regard. Future studies could explore visualizing and interpreting the learned behavior of the pruned model ensembles and their application to other screening situations like COVID-19 detection and localization in 3D CT scans, etc. At present, we expect that the proposed approach can be quickly adapted for detection of COVID-19 pneumonia using digitized chest radiographs.

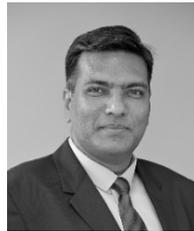

**SIVARAMAKRISHNAN RAJARAMAN** (Member, IEEE) received his Ph.D. in Information and Communication Engineering from Anna University, India. He is involved in projects that aim to apply computational sciences and engineering techniques toward advancing life science applications. These projects involve the use of medical images for aiding healthcare professionals in low-cost decision-making at the point of care screening/diagnostics. He is a versatile researcher with expertise in machine learning, data science, biomedical image analysis/understanding, and computer vision. He is a Life Member of the International Society of Photonics and Optics, a regular member of the Institute of Electrical and Electronics Engineers (IEEE), the IEEE Engineering in Medicine and Biology Society, and Biomedical Engineering Society (BMES).

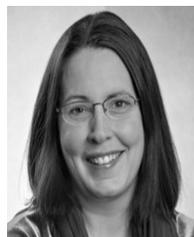

**JENIFER SIEGELMAN** serves as the Senior Medical Director of Takeda Pharmaceuticals, Cambridge, MA. She received her Doctor of Medicine (M.D.) from the State University of New York, Downstate Medical Center College of Medicine, and Master of Public Health (MPH) from Johns Hopkins Bloomberg School of Public Health. She served as the Radiology Resident/Fellow of Body Imaging at the Johns Hopkins School of Medicine.

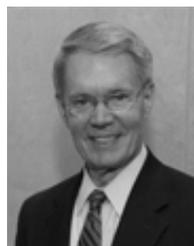

**PHILIP. O. ALDERSON** served as Dean of Saint Louis University School of Medicine from 2008 through 2016 and as Vice-President for Medical Affairs from 2009 through 2016. He was formerly the James Picker Professor and Chairman of the Department of Radiology at the Columbia University College of Physicians and Surgeons, Radiologist-in-Chief at New York-Presbyterian Hospital/Columbia University Medical Center and President of the Medical Board at New York-Presbyterian. His current interests include health care reform, big data analytics, genomic medicine, and structural biology. He is former President of the Society of Chairmen of Academic Radiology Departments, the Association of University Radiologists, the Association of Residency Program Directors in Radiology, the Academy of Radiology Research, the Fleischner Society, the American Roentgen Ray Society and the American Board of Radiology. He is a Fellow of the American College of Radiology, the American Association for the Advancement of Science and the American Institute for Medical and Biomedical Engineering.

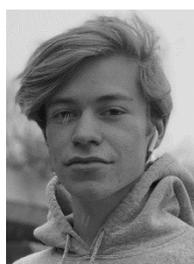

**LUCAS S. FOLIO** is doing his High-school diploma in computer science, at the Walt Whitman High School, Bethesda, MD. He is also working as a Bioinformatician at the NIH Clinical Center (CC). He served as a Web designer at RadSite where he performed Web development and hosting tasks using WordPress, HTML, and CSS tools. He is currently working on projects that involve web development, machine learning, information systems, and cybersecurity.

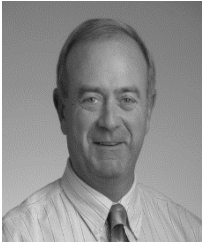

**LES R. FOLIO** is the lead radiologist for Computed Tomography (CT) in the Radiology and Imaging Sciences department at the NIH Clinical Center (CC). He is a Fellow of both the American Osteopathic College of Radiology (AOCR) and the Society of Computed Body Tomography and Magnetic Resonance. Dr. Folio was selected for one of the highest awards (the Trenary Medal) bestowed by AOCR. He has over 100 peer-reviewed publications and has authored five books. He is a co-author of several publications related to automated tuberculosis detection already deployed in Kenya and quantitative imaging in non-tuberculous mycobacteria.

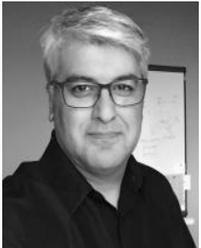

**SAMEER K. ANTANI** (Senior Member, IEEE) is a versatile lead researcher advancing the role of computational sciences and automated decision making in biomedical research, education, and clinical care. His research interests include topics in medical imaging and informatics, machine learning, data science, artificial intelligence, and global health. He applies his expertise in machine learning, biomedical image informatics, automatic medical image interpretation, data science, information retrieval, computer vision, and related topics in computer science and engineering technology His primary R&D areas include cervical cancer, HIV/TB, and visual information retrieval, among others. He is a Senior Member of the International Society of Photonics and Optics, Institute of Electrical and Electronics Engineers (IEEE) and the IEEE Computer Society.